\documentclass[8.5pt,twoside,twocolumn]{article}
\usepackage[super,sort&compress,comma]{natbib}
\usepackage{mhchem}
\usepackage{times,mathptmx}
\usepackage{sectsty}
\usepackage{balance}

\usepackage{graphicx} 
\usepackage{lastpage}
\usepackage[format=plain,justification=raggedright,singlelinecheck=false,font=small,labelfont=bf,labelsep=space]{caption}
\usepackage{fancyhdr}
\newcommand{\beq}{\begin{equation}}
\newcommand{\eeq}{\end{equation}}
\newcommand{\bea}{\begin{eqnarray}}
\newcommand{\eea}{\end{eqnarray}}

\newcommand{\benn}{\begin{displaymath}}
\newcommand{\eenn}{\end{displaymath}}

\newcommand{\Tup}{T_{\uparrow}}
\newcommand{\Tdn}{T_{\downarrow}}
\newcommand{\Sup}{S_{\uparrow}}
\newcommand{\Sdn}{S_{\downarrow}}
\newcommand{\Gup}{G_{\uparrow}}
\newcommand{\Gdn}{G_{\downarrow}}

\begin{document}

\thispagestyle{plain}
\fancypagestyle{plain}{
\fancyhead[L]{\includegraphics[height=8pt]{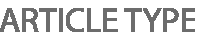}}
\fancyhead[C]{\hspace{-1cm}\includegraphics[height=20pt]{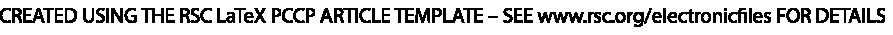}}
\fancyhead[R]{\includegraphics[height=10pt]{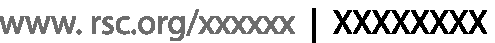}\vspace{-0.2cm}}
\renewcommand{\headrulewidth}{1pt}}
\renewcommand{\thefootnote}{\fnsymbol{footnote}}
\renewcommand\footnoterule{\vspace*{1pt}%
\hrule width 3.4in height 0.4pt \vspace*{5pt}}
\setcounter{secnumdepth}{5}

\makeatletter
\def\subsubsection{\@startsection{subsubsection}{3}{10pt}{-1.25ex plus -1ex minus -.1ex}{0ex plus 0ex}{\normalsize\bf}}
\def\paragraph{\@startsection{paragraph}{4}{10pt}{-1.25ex plus -1ex minus -.1ex}{0ex plus 0ex}{\normalsize\textit}}
\renewcommand\@biblabel[1]{#1}
\renewcommand\@makefntext[1]%
{\noindent\makebox[0pt][r]{\@thefnmark\,}#1}
\makeatother
\renewcommand{\figurename}{\small{Fig.}~}
\sectionfont{\large}
\subsectionfont{\normalsize}

\fancyfoot{}
\fancyfoot[LO,RE]{\vspace{-7pt}\includegraphics[height=9pt]{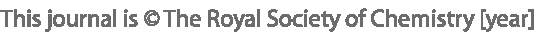}}
\fancyfoot[CO]{\vspace{-7.2pt}\hspace{12.2cm}\includegraphics{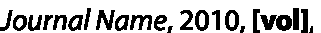}}
\fancyfoot[CE]{\vspace{-7.5pt}\hspace{-13.5cm}\includegraphics{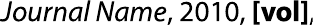}}
\fancyfoot[RO]{\footnotesize{\sffamily{1--\pageref{LastPage} ~\textbar  \hspace{2pt}\thepage}}}
\fancyfoot[LE]{\footnotesize{\sffamily{\thepage~\textbar\hspace{3.45cm} 1--\pageref{LastPage}}}}
\fancyhead{}
\renewcommand{\headrulewidth}{1pt}
\renewcommand{\footrulewidth}{1pt}
\setlength{\arrayrulewidth}{1pt}
\setlength{\columnsep}{6.5mm}
\setlength\bibsep{1pt}

\twocolumn[
  \begin{@twocolumnfalse}
\noindent\LARGE{\textbf{Enhanced thermoelectric efficiency in ferromagnetic silicene nanoribbons
asymmetrically terminated with hydrogen atoms}}
\vspace{0.6cm}

\noindent\large{\textbf{K. Zberecki$^1$,  R. Swirkowicz$^1$, J. Barna\'s$^2$}}\vspace{0.5cm}

\noindent\textit{\small{\textbf{Received Xth XXXXXXXXXX 20XX, Accepted Xth XXXXXXXXX 20XX\newline
First published on the web Xth XXXXXXXXXX 200X}}}

\noindent \textbf{\small{DOI: 10.1039/b000000x}}
\vspace{0.6cm}

\noindent \normalsize{Using ab-initio methods we calculate thermoelectric and spin thermoelectric
properties of silicene nanoribbons with bare, mono-hydrogenated and
di-hydrogenated edges. Asymmetric structures, in which one edge is either bare or di-hydrogenated while the
other edge is mono-hydrogenated (0H-1H and 2H-1H nanoribbons) have ferromagnetic ground state and
display remarkable conventional and spin thermoelectric  properties. Strong enhancement of the thermoelectric
efficiency, both conventional and spin ones, results from a very specific band
structure of such nanoribbons, where one spin channel is blocked due to an energy gap while the
other spin channel is highly conducting. In turn, 0H-2H and 2H-2H nanoribbons (with one edge being either bare or di-hydrogenated and the other edge being di-hydrogenated) are antiferromagnetic
in the ground state. Accordingly, the corresponding spin channels are equivalent, and only
conventional thermoelectric effects can occur in these nanoribbons.}
\vspace{0.5cm}
 \end{@twocolumnfalse}
  ]

\section{Introduction}


\footnotetext{\textit{$^{1}$Faculty of Physics, Warsaw University of Technology, ul. Koszykowa 75, 00-662 Warsaw, Poland}}
\footnotetext{\textit{$^{2}$~Faculty of Physics, Adam Mickiewicz University, ul. Umultowska 85, 61-614 Pozna\'n, Poland\\ and Institute of Molecular Physics, Polish Academy of Sciences, Smoluchowskiego 17, 60-179 Pozna\'n, Poland}}


Thermoelectric phenomena in various nanostructures are currently of great interest due
to the possibility of converting dissipated heat to electrical energy at
nanoscale.  Some nanostructures can exhibit relatively high thermoelectric
efficiency, being thus suitable candidates for applications in nanoelectronics
devices as elements of power generators or cooling systems. It is already well
known that quantum confinement and Coulomb blockade in nanoscale junctions with
quantum dots and molecules can lead to a considerable enhancement of the heat to
current conversion efficiency~\cite{b1,b2,b3}.

Very recently, new
two-dimensional materials like graphene and silicene -- especially
one-dimensional graphene (GNRs) and silicene (SiNRs) nanoribbons -- have been attracting
a great interest due to their unique  properties~\cite{b4,b5,b6,b7,b8}. Though
thermopower $S$ of pristine graphene is not very high ($S$ close to 100
$\mu$V/K has been reported~\cite{b9}) its value can be considerably enhanced in
GNRs, especially in nanostructures consisting of nanoribbons of various types.
Indeed, in a properly designed nanoribbon with alternating zigzag and armchair
sections, thermoelectric figure of merit exceeding unity at room temperature has
been found~\cite{b10}. The efficiency can be also enhanced by
randomly distributed hydrogen vacancies in almost completely hydrogenated
GNRs~\cite{b11}. Furthermore, structural defects, especially in the form of antidots, also
appear a promising way to enhance thermoelectric efficiency~\cite{b12,b13,b14}.
Indeed, giant spin related
thermoelectric phenomena have been predicted for ferromagnetic zGNRs with antidots~\cite{b14}.

In nanoscale systems, in which two spin channels are not equivalent and not
mixed by spin-flip processes, spin related thermoelectric phenomena can be
observed in addition to the conventional thermoelectric effects. These spin
related phenomena result from interplay of spin, charge and heat
transport~\cite{b15}. For instance, in ferromagnetic nanostructures one can observe the spin Seebeck effect,
which is a spin analog of the well known Seebeck effect and is associated with thermal generation of spin voltage. This phenomenon was
observed in thin films~\cite{b16,b17} and also in magnetic tunnel
junctions~\cite{b18,b19}.

Since silicon plays a crucial role in the present-day electronics, integration
of silicene -- a two-dimensional hexagonal lattice of silicon atoms -- into
nanoelectronics seems to be more promising than that of graphene. Therefore,
understanding of physical properties of silicene is currently of great interest.
Electronic structure of silicene is similar to that of graphene, i.e. silicene
is a semimetal with low-energy states at the Fermi level described by the Dirac
model~\cite{b20}. However, due to the buckled atomic structure it is possible to
open an energy gap with electric field~\cite{b21,b22}. Quite recently, silicene
nanoribbons have been fabricated~\cite{b23}, and their properties have been
widely studied  theoretically by ab-initio
methods~\cite{b24,b25,b26,b27,b28,b29}. Silicene nanoribbons of zigzag type,
similarly to graphene ones, can exhibit magnetic ordering of edge moments.
Ab-initio calculations show that the lowest-energy state of nanoribbons with
edges symmetrically terminated with atomic hydrogen is antiferromagnetic, where
magnetic moments at one edge are antiparalell to those at the other
edge~\cite{b27,b28,b29}.  Such a  nanoribbon behaves like a semiconductor, with a
relatively wide energy gap. However, in the presence of external magnetic field,
a ferromagnetic state with all moments ordered in parallel can be
stabilized~\cite{b28}. The system behaves then as a metallic ferromagnet with
constant transmission in the vicinity of the Fermi level~\cite{b28}. On the
other hand, the one-side ferromagnetic state, with magnetic moments localized
only along one edge, can be induced in the presence of densely distributed
impurities (like boron or nitrogen) at the other edge~\cite{b30,b31}. The nanoribbon behaves then as a spin
gapless semiconductor, where one spin channel is blocked for conduction
due to an energy gap while the second spin channel is gapless.

Experimental and theoretical studies have shown that electronic and magnetic
properties of zSiNRs are strongly influenced by the presence of
hydrogen~\cite{b29,b32}. Based on the density-functional-theory (DFT)
calculations it has been found that at low hydrogen concentration,  zSiNRs nanoribbons
with edges terminated with atomic hydrogen are stable, whereas at ambient
conditions nanoribbons with dihydrogenated edges can appear~\cite{b29}. It has
been also predicted that hydrogen-terminated zSiNRs subject to in-plane
electrical field behave like a half-metallic ferromagnet with spin polarization
up to 99\%~\cite{b33}.

In the present paper we study the effects of edge hydrogenation on electronic
transport and thermoelectric properties of zSiNRs. The considerations clearly show that
nanoribbons with asymmetrically terminated edges, where one edge is
mono-hydrogenated while the other edge is either bare or dihydrogenated, exhibit stable ferromagnetic
state and also pronounced thermoelectric properties -- both conventional and spin
ones. The predicted thermoelectric efficiency is exceptionally high and results
from the very specific band structure, with the gaps corresponding to the two spin channels remarkably
shifted in energy. Accordingly, one spin channel can be conductive in a certain
region of chemical potential, whereas the second channel is then blocked for
transport and thus gives rise to a high thermopower. Due to their remarkable electronic and thermoelectric
properties, ferromagnetic zSiNRs with hydrogenated edges have a huge potential for
various applications in nanoelectronic and nanothermoelectric devices.

The paper is organized as follows. In section 2 we describe the computational method used to calculate band structure and transmission function.
We also provide there a general description of thermoelectric properties. Numerical results on electric and thermoelectric properties are
presented and discussed in section 3. Final conclusions are in section 4.

\section{Theoretical description}

\subsection{Transmission}

To investigate electronic transport through zSiNRs, we used the ab-initio approach
within the the DFT Siesta code~\cite{siesta1,siesta2}. Numerical calculations were
performed for nanoribbons containing N=6 zigzag chains. In order to analyze the influence
of hydrogen termination at the nanoribbon edges on the magnetic, transport and
thermoelectric properties, we will consider the following four different situations: (i) 1H-0H
systems, in which one edge of zSiNRs is terminated with atomic hydrogen, whereas
the dangling bonds appear at the other edge; (ii) 1H-2H structures, where one edge
is mono-hydrogenated and the other one is dihydrogenated; (iii) 2H-0H nanoribbons with
one edge dihydrogenated and the other being free of hydrogen; and (iv) 2H-2H
nanoribbons, with both edges being dihydrogenated. The simple structure, 1H-1H, with both edges
terminated with atomic hydrogen was discussed earlier~\cite{b28}, so
this situation will be omitted in the present paper.

All the structures under
consideration were optimized until atomic forces converged to 0.02 eV/A. Furthermore, the
atomic double-polarized basis (DZP) was used and the grid mesh cutoff was set
equal to 200 Ry. Apart from this, the generalized gradient approximation (GGA) with
Perdrew-Burke-Ernzerhof parameterization was applied to the exchange-correlation
part of the total energy functional~\cite{pbe1}.

\begin{figure}[ht]
  \begin{center}
    \begin{tabular}{cc}
      \resizebox{40mm}{!}{\includegraphics[angle=0]{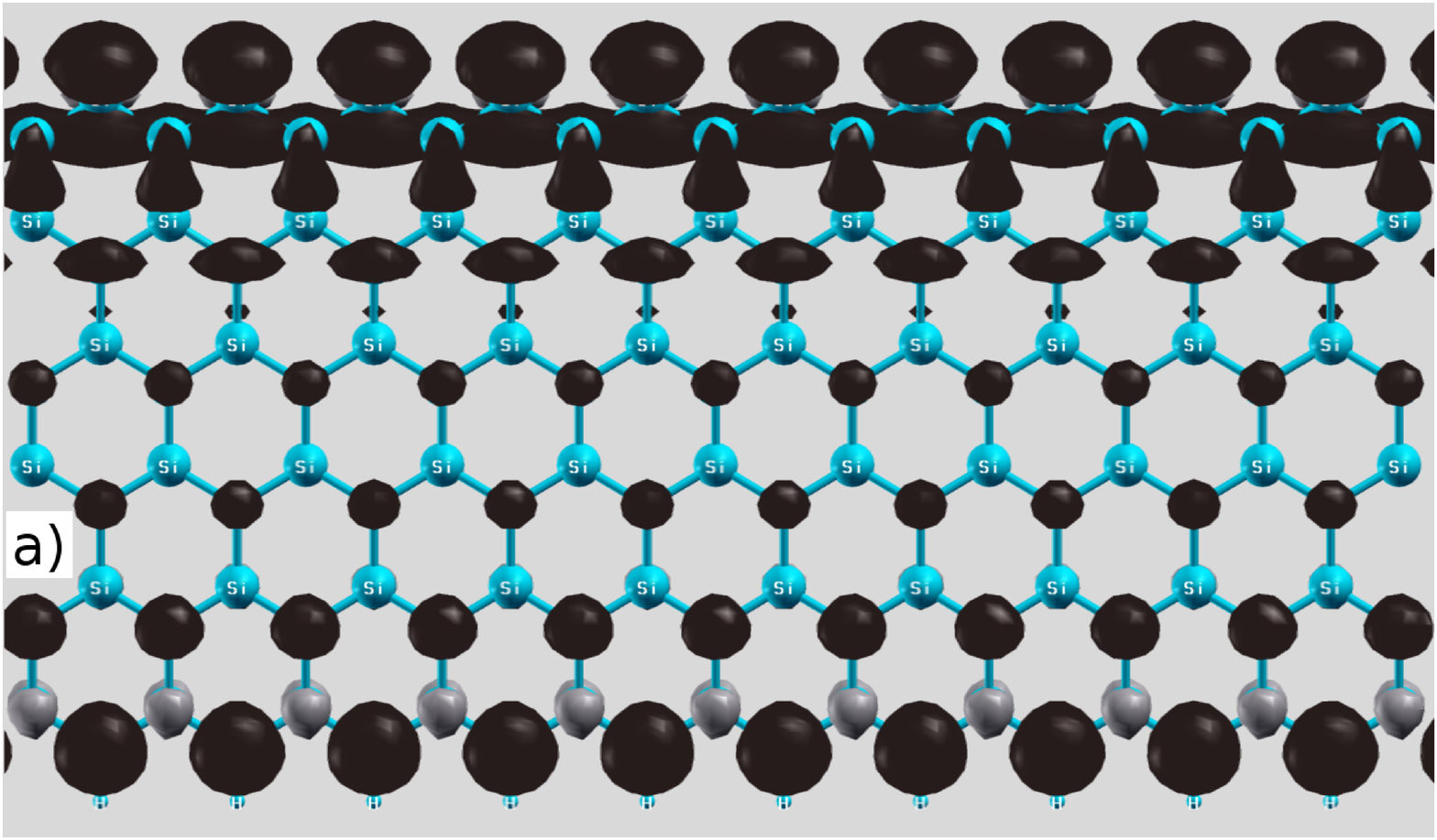}} &
      \resizebox{40mm}{!}{\includegraphics[angle=0]{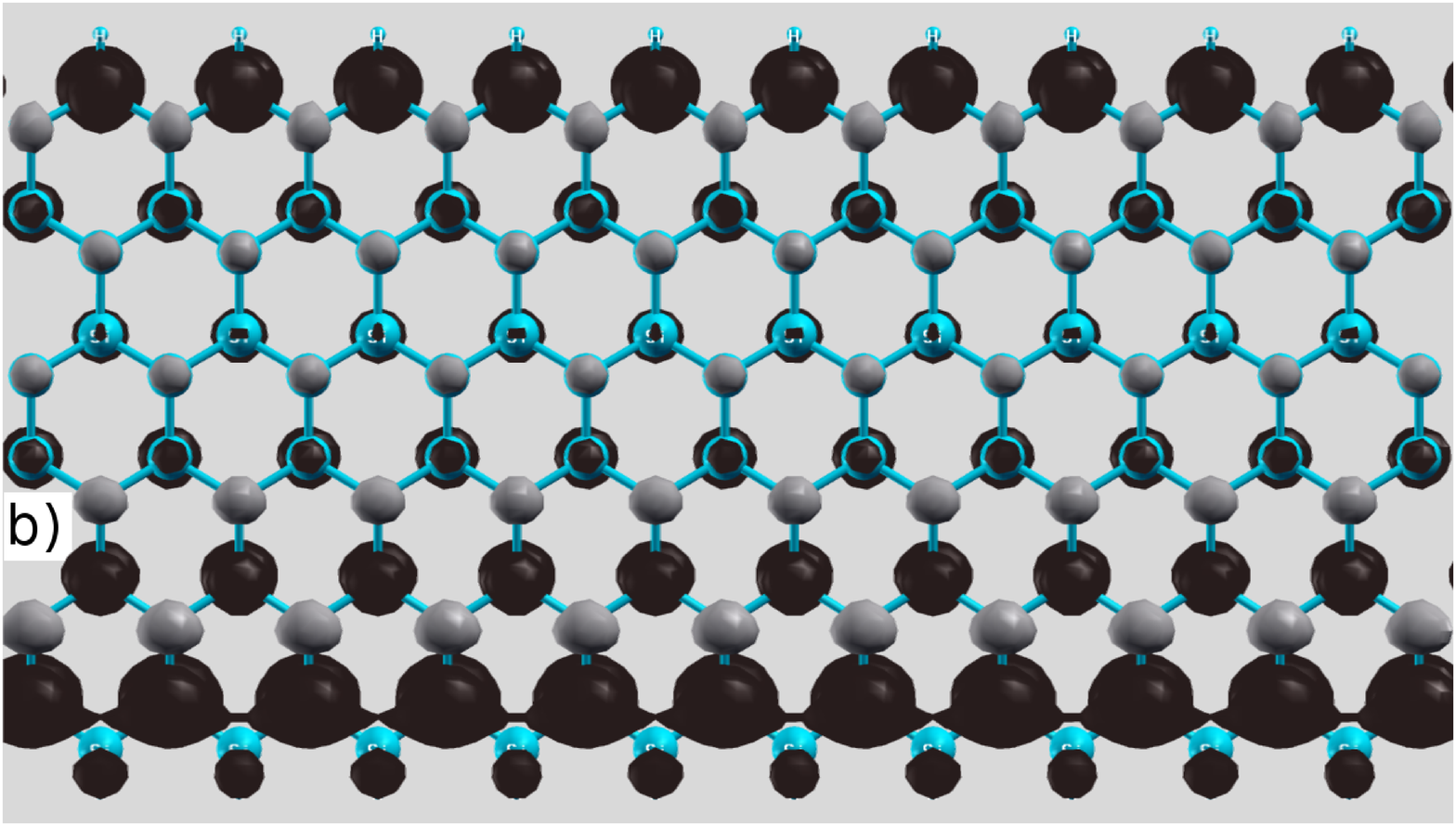}} \\
      \resizebox{40mm}{!}{\includegraphics[angle=0]{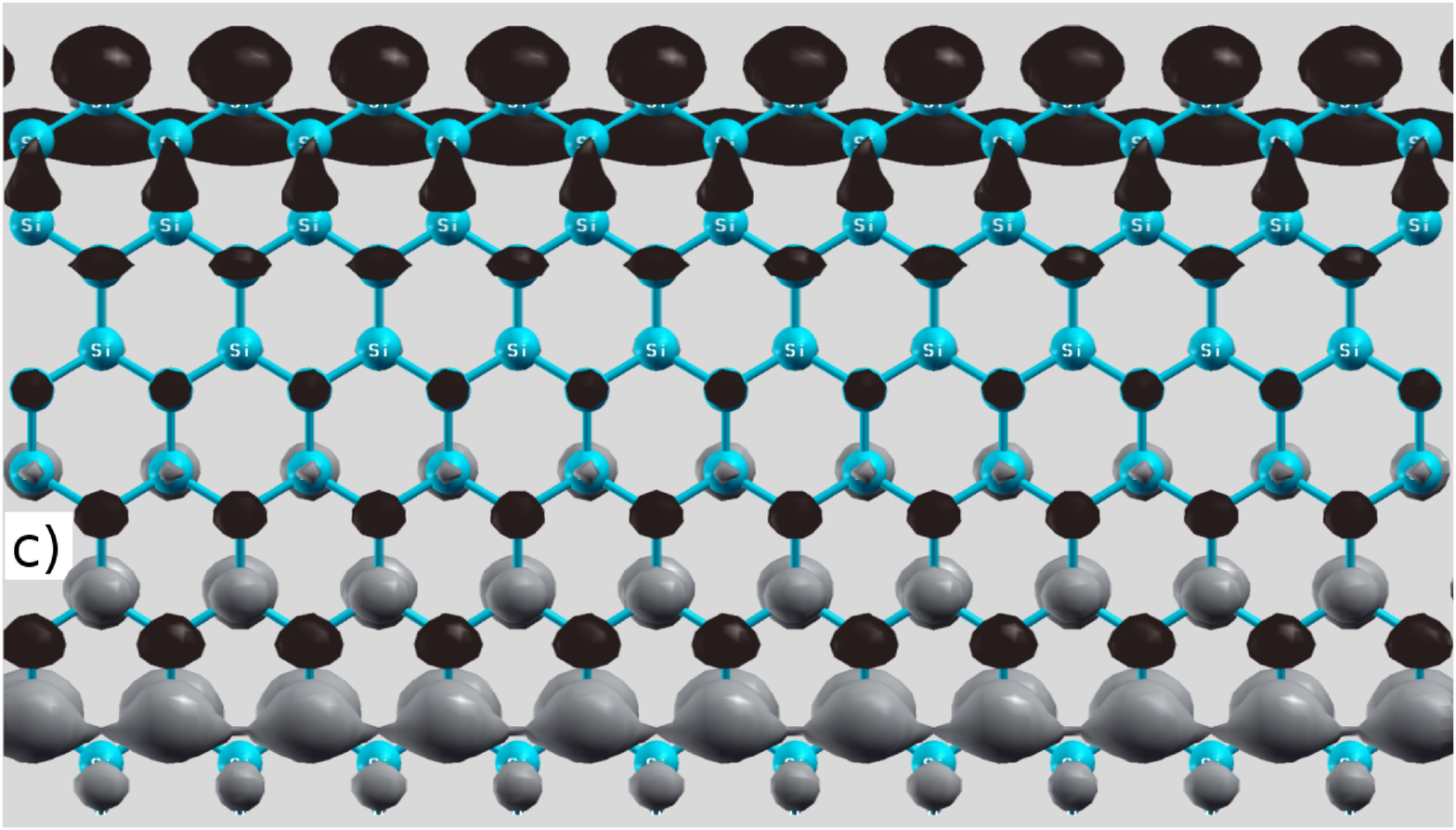}} &
      \resizebox{40mm}{!}{\includegraphics[angle=0]{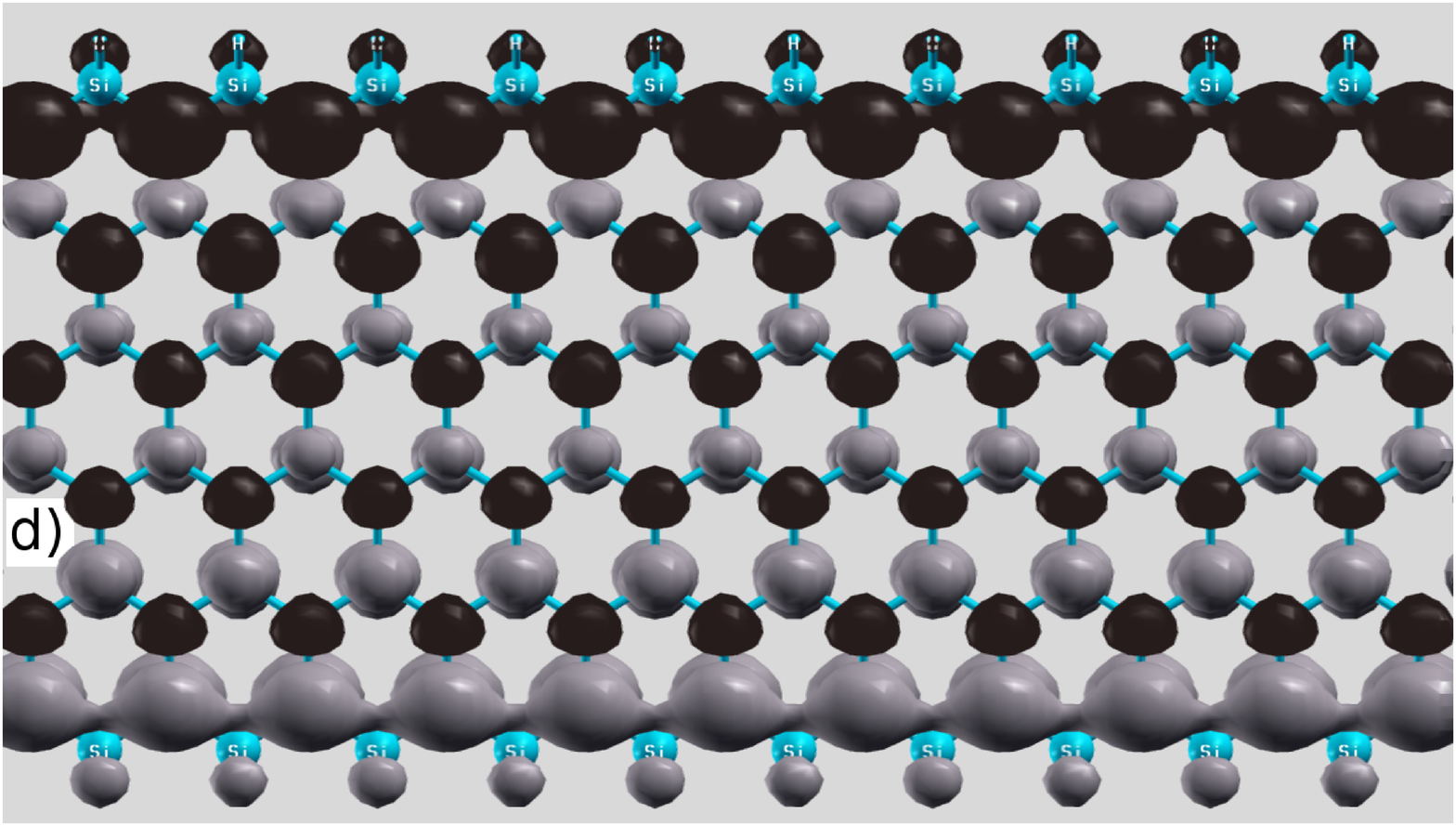}} \\
    \end{tabular}
    \caption{(Color online) Spin density distribution in the ferromagnetic (FM) (a,b) and antiferromagnetic (AFM) (c,d) ground states,
calculated for zSiNR of 1H-0H (a), 1H-2H (b), 2H-0H (c) and 2H-2H (d) types.
Black and grey dots represent magnetic moments of opposite orientations.}
    \label{fig1}
  \end{center}
\end{figure}

Numerical calculations show that the nanoribbons under consideration exhibit magnetic ordering
in the ground state. The corresponding spin density distribution across the
nanoribbons is presented in Fig. 1. The
ground state of zSiNRs with 1H-0H and 1H-2H edges is ferromagnetic (FM), with
the corresponding energy lower than that of antiferromagnetic (AFM) configuration  by 0.01 eV
and 0.02 eV, respectively. On the other hand, the most stable configuration of
the nanoribbons  of 2H-0H and 2H-2H types is antiferromagnetic, with magnetic
moments at one edge being opposite to those at the other edge (Figs 1c,d).
Energy of the AFM state is smaller than the energy of the corresponding FM state, respectively  by 0.06 eV
for 2H-0H  and 0.01 eV for 2H-2H nanoribbons. Other possible configurations, eg. one-sided
ferromagnetic or nonmagnetic ones, have considerably higher energy.

The spin-resolved energy-dependent transmission function $T_{\sigma}(E)$ through
the nanoribbons was determined within the non-equilibrium Green function method
(NGF) as implemented in the Transiesta code~\cite{siesta2}. Having determined
the transmission $T_{\sigma}(E)$, one  can calculate transport properties (including also
thermoelectric ones) of the nanoribbons, as described below.

\subsection{Thermoelectricity}

When the spin channels are mixed in the nanoribbon on a distance comparable to
the system length, no spin related thermoelectric properties can be observed and
only conventional thermoelectric phenomena occur. We will consider first this
limit.
In the linear response regime, the electric $I$ and heat $I_{Q}$ currents
flowing through the system from left to right, when the electrical potential and
temperature of the left electrode are higher by  $\delta V$ and $\delta T$,
respectively, can be written in the matrix form as~\cite{b28,sivan,Mahan}
\beq
\left(
\begin{array}{c}
I   \\
I_{Q} \\
\end{array}
\right)
=
\left(
\begin{array}{cc}
e^2 L^+_{0} & \frac{e}{T}L^+_{1} \\
 eL^+_{1} &  \frac{1}{T}L^+_{2} \\
\end{array}
\right)
\left(
\begin{array}{c}
\delta V   \\
\delta T \\
\end{array}
\right),
\eeq
where $e$ is the electron charge, while
$L^+_{n} = L_{n \uparrow} + L_{n \downarrow}$, with $L_{n \sigma} = -\frac{1}{h}
\int dE\,T_{\sigma}(E)\, (E-\mu)^{n} \frac{\partial f}{\partial E} $
for $n=0,1,2$. Here, $T_{\sigma}$(E) is the spin-dependent transmission function
(described above) for the system and
$f(E-\mu)$ is the Fermi-Dirac distribution function corresponding to the
chemical potential $\mu$
and temperature $T$. The electrical conductance $G$ (for $\delta T=0$) is given as $G=e^{2}
(L_{0\uparrow}+L_{0\downarrow})\equiv G_\uparrow+G_\downarrow$, whereas the
electronic contribution to the thermal conductance, $\kappa_{e}$, defined by heat
current at $I=0$, is equal to
\begin{equation}
\kappa_{e}=\frac{1}{T} \left(L^+_{2} - \frac{L_{1}^{+2}}{L^+_{0}}\right).
\end{equation}
The thermopower, defined as $S=-\delta V /\delta T$ for  $I=0$, is given by the
formula
\begin{equation}
S=-\frac{L^+_{1}}{|e|TL^+_{0}}.
\end{equation}
In turn, the thermoelectric efficiency of the system is described by the figure
of merit $ZT$ defined as
\begin{equation}
ZT = \frac{S^{2}GT}{\kappa},
\end{equation}
where $\kappa$ corresponds to the total thermal conductance due to electrons and
phonons,  $\kappa = \kappa_{e} + \kappa_{ph}$. \newline

Situation may change when the spin channels are either not mixed in the
nanoribbon or they are mixed on a scale much longer than the system's size.
The spin effects in thermoelectricity become then important and to describe them
one needs to introduce spin voltage $\delta V_s$ and also spin current $I_s$.
The formula (1) can be then generalized to the following one:
\beq
\left(
\begin{array}{c}
I   \\
I_{s} \\
I_Q\\
\end{array}
\right)
=
\left(
\begin{array}{ccc}
e^2 L^+_{0} & e^2L^-_{0} & \frac{e}{T}L^+_{1} \\
e^2 L^-_{0} & e^2L^+_{0} & \frac{e}{T}L^-_{1} \\
 eL^+_{1} & eL^-_{1}  &  \frac{1}{T}L^+_{2} \\
\end{array}
\right)
\left(
\begin{array}{c}
\delta V   \\
\delta V_s   \\
\delta T \\
\end{array}
\right),
\eeq
where the spin current $I_s$ is normalized to $\hbar /2e$, while $L^-_{n} = L_{n \uparrow} - L_{n \downarrow}$ and $L_n^+$ is defined as above.
The electrical conductance (for $\delta V_s=0$ and $\delta T=0$) of the
system is then given by the formula
$G = e^{2}(L_{0\uparrow}+L_{0\downarrow})\equiv G_\uparrow+G_\downarrow$, while
the spin conductance (for $\delta V=0$ and $\delta T=0$) is given by
$G_{s} = e^{2}(L_{0\uparrow}-L_{0\downarrow}) =  G_\uparrow -G_\downarrow$. In
turn,
the electronic contribution to the thermal conductance, $\kappa_{e}$, defined by the heat current at
$I=0$ and $I_s=0$, is given by~\cite{b30}
\begin{equation}
\kappa_{e}=\frac{1}{T} \sum_{\sigma} \left(L_{2\sigma} -
\frac{L_{1\sigma}^{2}}{L_{0\sigma}}\right).
\end{equation}
The conventional (charge) thermopower can be defined as $S_c=-\delta V/\delta T$
at $I=0$ and $I_s=0$, while the spin thermopower as
$S_s=-\delta V_s/\delta T$ also for $I=0$ and $I_s=0$. Thus, the thermopowers
can be written in the form~\cite{b30}
\begin{eqnarray}
S_c=-\frac{1}{2|e|T}(L_{1\uparrow} / L_{0\uparrow} + L_{1\downarrow} /
L_{0\downarrow})\equiv \frac{1}{2}(S_{c\uparrow} + S_{c\downarrow} ), \nonumber
\\
S_s=-\frac{1}{2|e|T}(L_{1\uparrow} / L_{0\uparrow} - L_{1\downarrow} /
L_{0\downarrow})\equiv \frac{1}{2}(S_{c\uparrow} - S_{c\downarrow} ).
\end{eqnarray}
To describe the conventional (charge) and spin thermoelectric
efficiency, one can introduce the corresponding figures of merit, defined as
\beq
ZT_{c} = \frac{S_{c}^{2}G T}{\kappa}, \ \
ZT_{s} = \frac{S_{s}^{2}|G_s|T}{\kappa}.
\eeq

Alternatively, one can define the conventional thermopower as
$S_c=-(\delta V/\delta T)$ for $I=0$ and $\delta V_s=0$, and the spin
thermopower as
$S_s=-(\delta V_s/\delta T)$ for $I_s=0$ and $\delta V=0$~\cite{dubi09}. The thermopowers are then given by the formulas
$S_c=-L^+_{1}/(|e|TL^+_{0})$ and $S_s=-L^-_{1}/(|e|TL^+_{0})$.
Similarly, the heat conductance can be defined by the heat current at $I=0$ and
$\delta V_s=0$, and thus is given by Eq.(6). The two different definitions
depend essentially on experimental conditions. In the following numerical
calculations we will use the former definitions given by Eqs (6-8).

In the linear response regime, considered here, transport properties are
determined by electronic states near the Fermi level. In reality the chemical
potential in nanoribbons (measured from the Fermi energy)  can be easily varied
near the Fermi level with an external gate voltage. Alternatively, the chemical
potential can be moved down or up by p-type or n-type doping, which results in
$\mu<$0 and $\mu>$0, respectively.

\section{Ferromagnetic zSiNRs}

In nanoribbons with FM ordering of the edge magnetic moments, the  spin-up and
spin-down channels are generally different. If the corresponding spin
relaxation time is long enough, the spin thermoelectric effects become observable.
Assuming this is the case, we consider spin-dependent transport and spin
thermoelectric phenomena in zSiNRs of 1H-0H and 1H-2H types ~\cite{b34}. The limit of
independent of spin transport, when  only conventional thermoelectric effects
can be observed, will be considered in the next section for systems with AFM ordering.

\subsection{zSiNRs of 1H-0H type}

Spin-polarized electronic bands as well as the spin-resolved transmission function, calculated for the
ferromagnetic 1H-0H nanoribbon, are presented in Fig.2.
\begin{figure}[ht]
    \begin{tabular}{c}
      \resizebox{85mm}{!}{\includegraphics[angle=270]{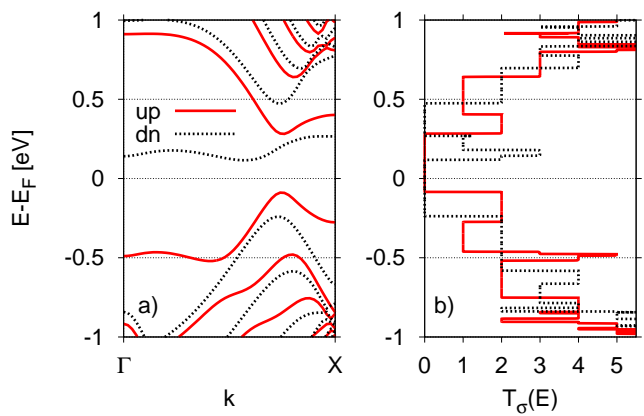}} \\
    \end{tabular}
    \caption{(Color online) Spin-resolved band structure (a) and transmission
function (b) calculated for zSiNR of 1H-0H type. The energy is measured with respect
to the Fermi energy $E_{F}$ of the corresponding pristine nanoribbon.}
    \label{fig2}
\end{figure}
This figure clearly shows, that the nanoribbon behaves like a ferromagnetic semiconductor with a
spin-dependent energy gap in the vicinity of the Fermi level. When the energy $E$ decreases below the Fermi energy $E_F$, then the first electronic states   below the
Fermi level correspond to the spin-up channel and are located in the spin-down
gap. The situation is different for energies above $E_{F}$, where a narrow  spin-down
band is located in the spin-up gap, but close to the gap edge. With a further increase in energy, another
gap for spin-minority (spin-down) electrons opens in the spin-up band. Thus, there is a spin dependent energy gap at the Fermi level,
and one additional gap for spin-down electrons at higher energies. All these gaps are very well visible in the
spin-resolved transmission function presented in Fig.2b. Moreover, these gaps strongly
influence spin-dependent transport properties, including also thermoelectric phenomena, as shown below for the temperature of 90 K.

\begin{figure*}[ht]
  \begin{center}
    \begin{tabular}{cc}
      \resizebox{75mm}{!}{\includegraphics[angle=270]{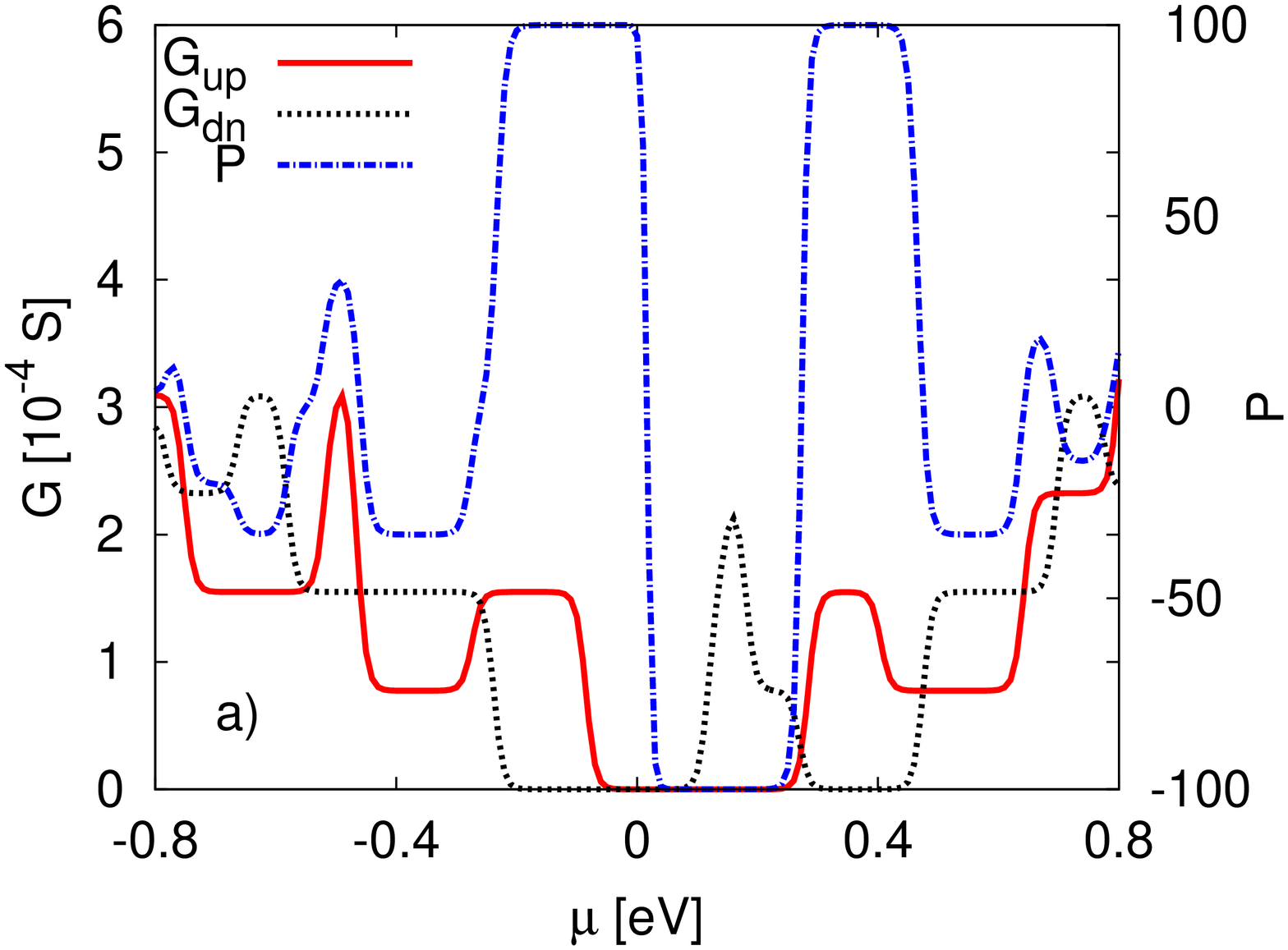}} &
      \resizebox{75mm}{!}{\includegraphics[angle=270]{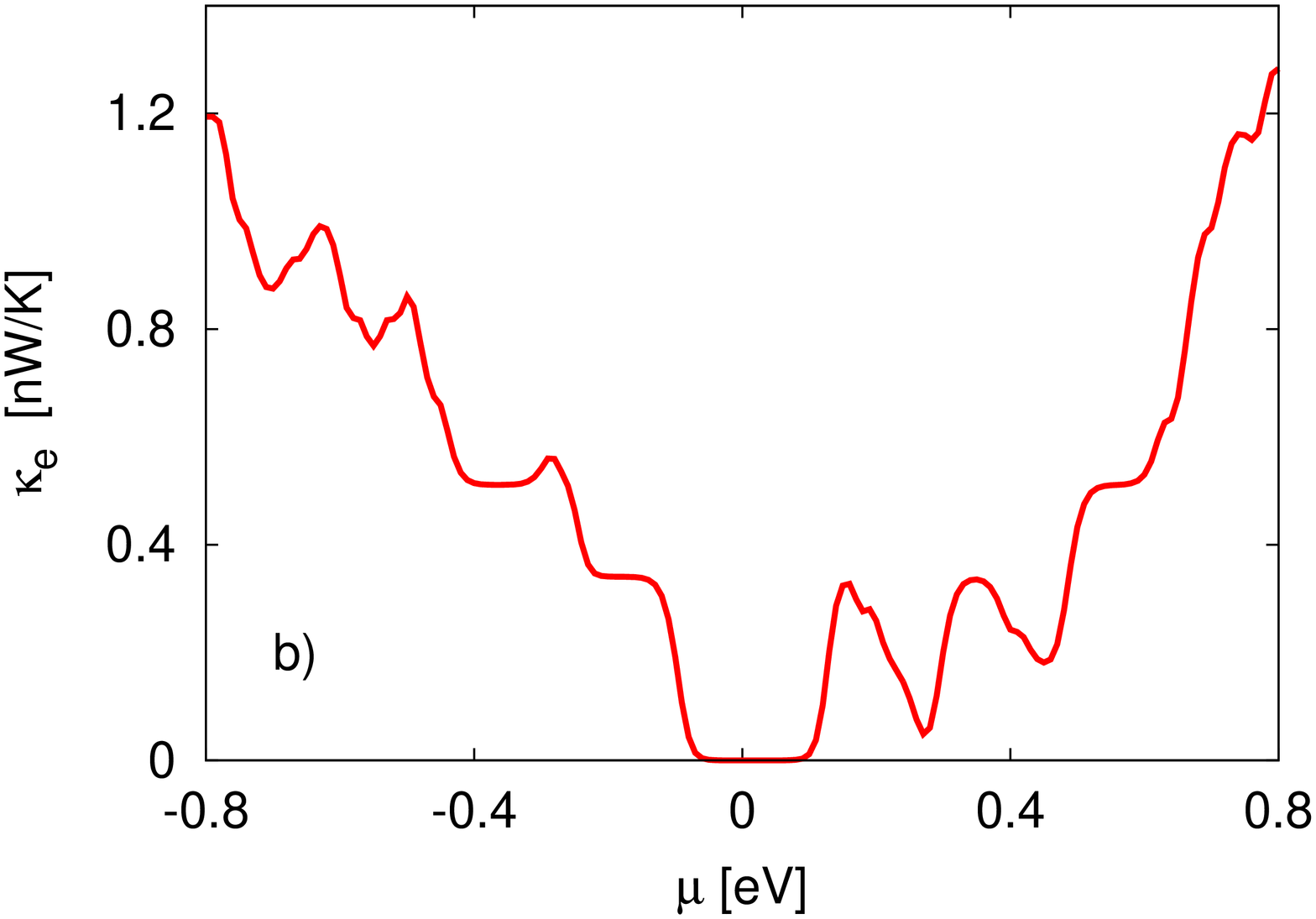}} \\
      \resizebox{75mm}{!}{\includegraphics[angle=270]{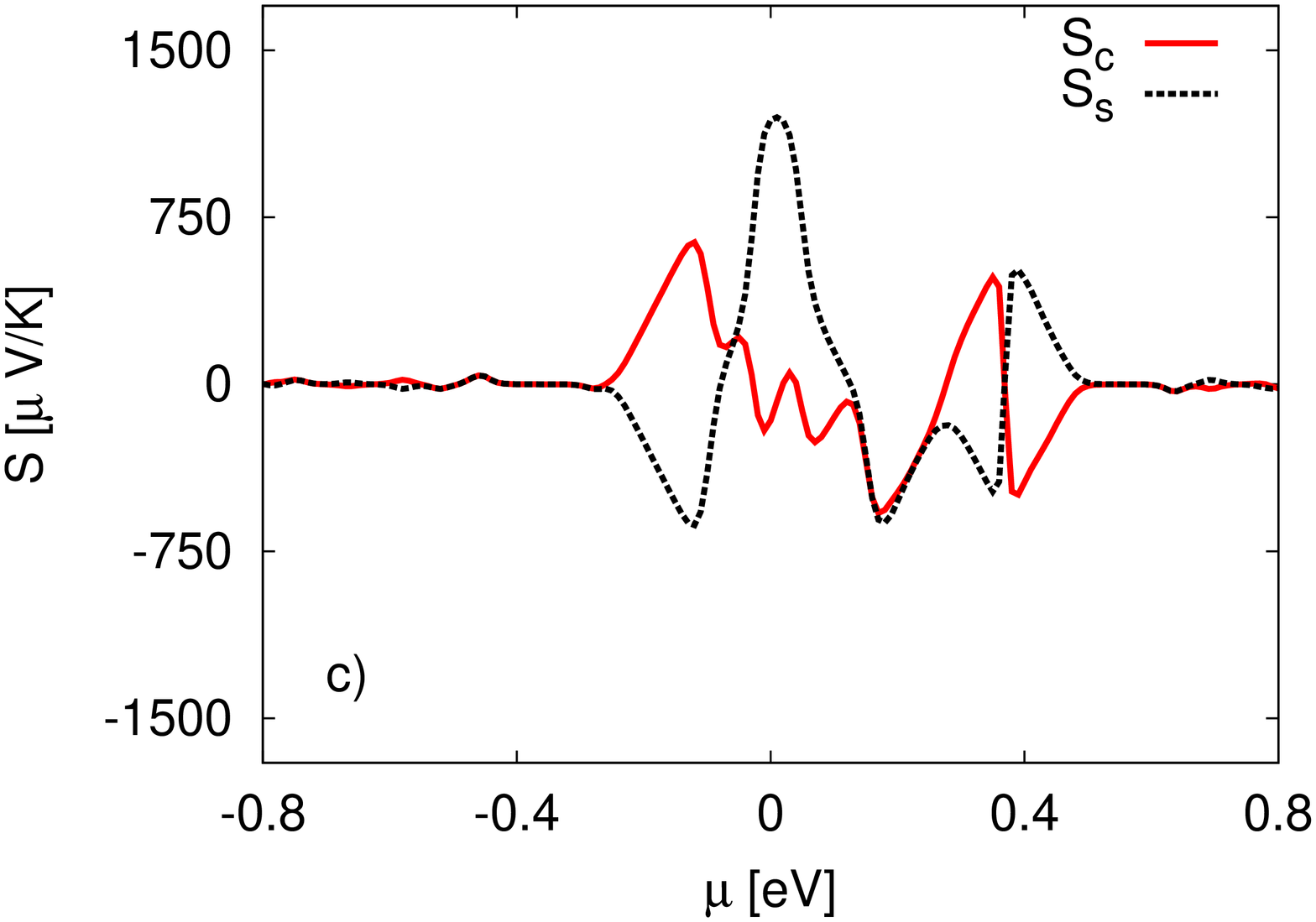}} &
      \resizebox{75mm}{!}{\includegraphics[angle=270]{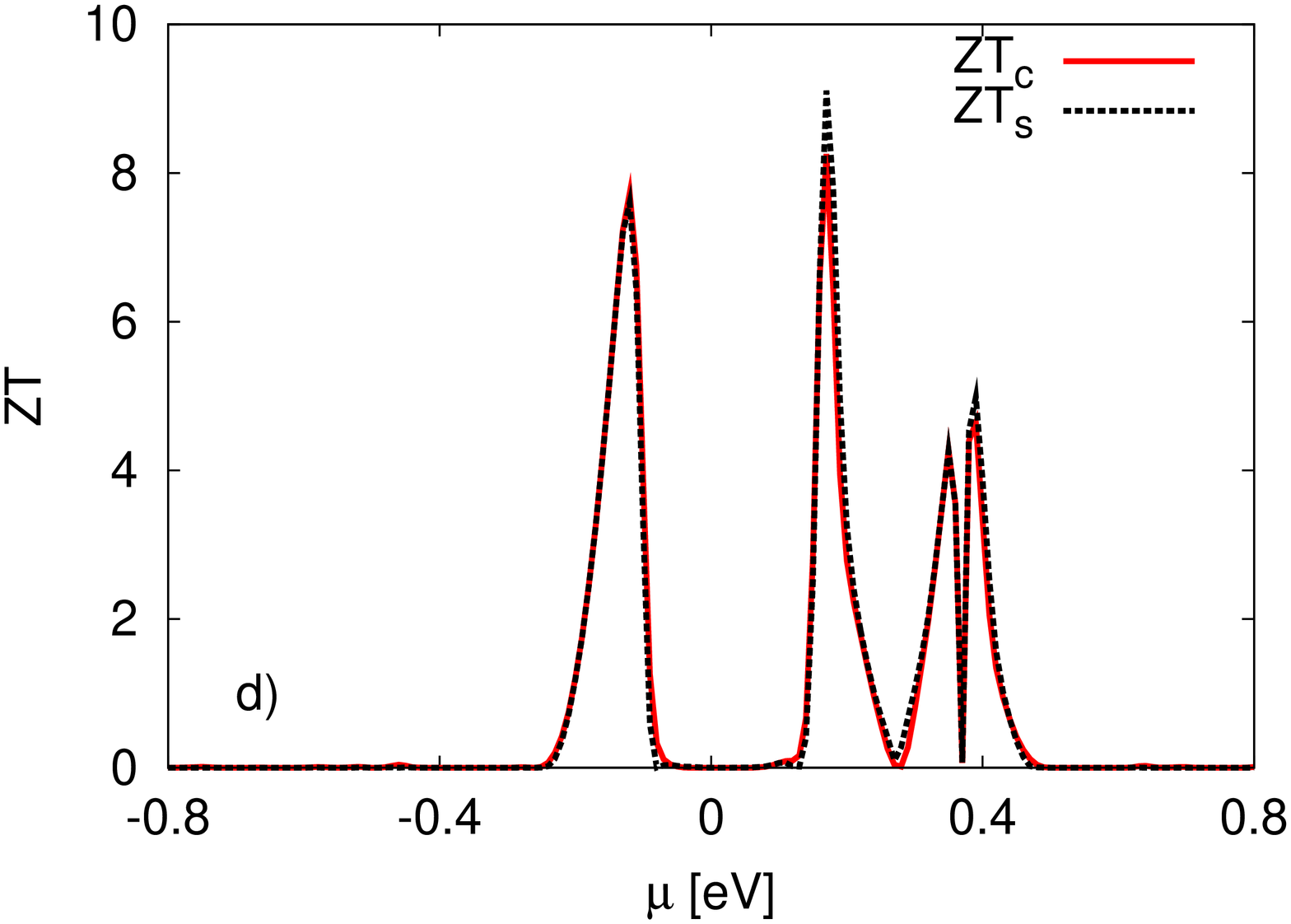}} \\
    \end{tabular}
    \caption{(Color online) Spin-resolved electric conductance $G_\sigma$ ($G_{up}$ for spin-up and $G_{dn}$ for spin-down electrons) 
    and the corresponding spin
polarization $P$ (a); electronic
term in the thermal conductance, $\kappa_{e}$ (b); charge, S$_{c}$, and spin, S$_{s}$,
thermopowers (c); and charge, ZT$_{c}$, and spin, ZT$_{s}$, figures of merit
(d); calculated for the zSiNR of 1H-0H type and $T=90$ K}
    \label{fig3}
  \end{center}
\end{figure*}

Spin resolved electrical conductance is presented in Fig. 3a as a function of the chemical potential
measured from the Fermi energy $E_F$. Due to the presence of spin-dependent energy gaps, there are regions of chemical
potential, where one of the spin channels is blocked for transport while the
other one is conductive. This leads to large spin polarization $P$ of the charge
current, defined as
$P=\frac{G_{\uparrow}-G_{\downarrow}}{G_{\uparrow}+G_{\downarrow}}\times 100\%$.
Note that for negative $\mu$, just below the Fermi level, only the spin-up
channel in conductive, whereas in a wide region of positive $\mu$ this channel
is blocked and the spin-down channel dominates transport. Accordingly, the
current polarization achieves $\pm 100 \%$, and changes  sign in the vicinity of
the Fermi level (corresponding to $\mu$=0). It is also interesting to note, that the electrical
conductance exhibits a considerably high maximum just above the Fermi level,
which appears  due to the presence of a relatively flat band corresponding to the
spin-down electrons. Thus, relatively small changes in the chemical potential allow 
to achieve quite good conduction, with almost total, $\pm 100\%$, spin
polarization. The situation is totally different in pristine zSiNRs
symmetrically terminated with atomic hydrogen, which exhibit AFM ordering with a
relatively wide energy gap. However, in the presence of external
magnetic field, such nanoribbons behave like a metallic ferromagnet with both spin
channels being conductive~\cite{b28}.

Thermal conductance due to electrons, $\kappa_{e}$, is presented in Fig. 3b as a function of the chemical potential $\mu$. 
Similarly to the electric conductance,
the thermal conductance is strongly suppressed  in the energy gap and remarkably increases
near the gap edges. This increase, however, is smaller than in zSiNRs symmetrically
terminated with atomic hydrogen, as only one spin channel, corresponding either to
spin-up or to spin-down electrons, can support heat current in the vicinity of
the gap edges.

Since the transmission through the system is strongly spin-dependent, one can
expect considerable spin effects in thermopower. The conventional $S_{c}$
and spin $S_{s}$ thermopowers are presented in Fig.3c as a function of
chemical potential. Indeed, a huge value of $S_{s}$, close to 1.4 mV/K, is obtained in
the vicinity of the Fermi energy (corresponding to $\mu=0$). Then, $S_{s}$ rapidly
decreases to zero with increasing $|\mu |$, changes sign, and its magnitude achieves values close to -0.7
mV/K for certain narrow regions of the chemical potential. For positive $\mu$,
 $S_{s}$ changes sign once more and achieves a local maximum of  $\approx 0.7$ mV/K.
For, larger values of $|\mu|$, when the chemical potential is well inside the electron bands and far from the gaps, the thermopower
 drops to small values, which are close to zero.
The conventional thermopower
$S_{c}$ is less remarkable, though for certain regions of $\mu$ it also takes
values close to 0.7 mV/K. High values of $S_{s}/S_{c}$ are related to the
presence of spin-dependent energy gaps in the spectrum, and especially to the very sharp transmission
changes near the gap edges. For small negative chemical potentials, transport
can be supported by spin-up holes as the spin-down channel is non-active due to
the energy gap in this channel. Then, the rapid decrease in $\Tup$
near E$_{F}$ generates a high positive contribution to $\Sup$. The situation is
different just above the Fermi level, where the spin-down channel becomes
conductive and the rapid increase in $\Tdn$ gives rise to a high negative contribution to
$\Sdn$. Accordingly, the thermopowers $\Sup$ and  $\Sdn$, corresponding to the
two spin channels, have opposite signs in the region close to $\mu$=0, and
therefore $S_{c}$ is small, while $S_{s}$ achieves the main maximum. When $|\mu|$ 
increases further, both $\Sup$ and $\Sdn$   decrease and become equal to zero when flat
parts of $\Tup$ or $\Tdn$ are achieved. For higher values of negative  $\mu$, the main
contribution comes then from the spin-down channel and corresponds to
a rapid decrease in $\Tdn$ at the edge of the gap in this spin channel. Both
$S_{c}$ and $S_{s}$ have  then opposite signs. Behavior of the thermopowers for positive $\mu$ is then slightly different due to the additional
gap in the spin-down channel as described above. This leads to additional positive local maximum in $S_s$ and local minimum in $S_c$
around $\mu=0.4$ eV, which appear close to the upper edge of the additional gap in the spin-down channel. Lower edge of this gap also leads
to some features in the thermopowers as clearly seen in Fig.3c.

Consider now the thermoelectric efficiency described by the figures of
merit $ZT_{c}$ and $ZT_{s}$. Both $ZT_{c}$ and $ZT_{s}$ are presented in Fig.3d as a
function of the chemical potential $\mu$. The thermoelectric efficiency was determined
here taking into account phonon contribution to the heat conductance, calculated for zSiNRs in our
earlier paper~\cite{b28}. Accordingly, the total heat conductance is a sum of the
electronic contribution $\kappa_{e}$, presented in Fig. 3b, and the phonon part
$\kappa_{ph}$ taken equal to 0.42 nW/K~\cite{b28}. Variation of both conventional $ZT_{c}$ and spin
$ZT_{s}$ figures of merit with the chemical potential reveals  two high  peaks located in the vicinity of $\mu$=0;
one for negative and another one for positive $\mu$. Apart from this, there are two additional
peaks of lower intensity in both $ZT_{c}$ and $ZT_{s}$, which appear for higher values of positive chemical
potential.

The thermoelectric efficiency calculated here for 1H-0H zSiNRs is very
remarkable, despite the total heat conductance is relatively high. Note that
nanoribbons symmetrically terminated with atomic hydrogen exhibit much smaller
thermoelectric efficiency~\cite{b28}. This huge enhancement of the figures of merit
comes mainly from the high power factors
$S_{c}^{2} G$ and $S_{s}^{2} G_s$.
Furthermore, $S_{c}$ and $S_{s}$ are strongly enhanced for $\mu<$0 and $\mu>$0
corresponding to the edges of energy gaps in the spin-down ($\mu<$0) and spin-up
($\mu>$0) channels. Additionally, the electrical conductance $\Gup$ for $\mu<$0
and $\Gdn$ for $\mu>$0 is relatively high as the appropriate channel is
conductive. Both power factors are thus enhanced, accordingly. Note, that the
efficiency $ZT_{s}$ shows no central peak, though the thermopower $S_{s}$ has a maximum 
for $\mu \approx 0$. However, both spin channels are blocked in this
region due to the energy gap in the spin-up and spin-down channels.
High efficiency can be observed only in situation when one spin channel is
blocked, while the second one is well conductive. It is worth of note, that similar enhancement of
the thermoelectric efficiency was found in ferromagnetic graphene nanoribbons
with structural defects in the form of antidots~\cite{b14}.

\subsection{2H-1H ferromagnetic nanoribbons}

Consider now the nanoribbons of 1H-2H type, where one edge is
mono-hydrogenated while the other edge is dihydrogenated. The lowest-energy
state of such nanoribbons is ferromagnetic, and the corresponding band structure is shown in Fig.
4a. In turn, Fig. 4b presents the  spin-resolved transmission
function. As follows from the band structure,
the system is a semiconductor with wide gaps at the Fermi level for both spin-up and spin-down channels.
These gaps, however, are shifted in energy, so the true energy gap at the Fermi level is rather narrow, despite
the gaps for individual spin channels are relatively wide, much wider than in the case of 1H-0H
nanoribbons analyzed  above.
However, there is now additional gap for spin-up channel (negative $\mu$) and also for spin-down channel (positive $\mu$).
Interestingly, the relatively flat spin-down band
above the Fermi energy is located  almost in the middle of the spin-up gap at the Fermi level. Thus, the
band structure in transmission is slightly reacher than in the case of  1H-0H
nanoribbons studied above, as there is one additional gap. This specific form of the transmission function 
makes the corresponding variation of the
thermoelectric parameters with the chemical potential more complex, as discussed below.
\begin{figure}[ht]
    \begin{tabular}{c}
      \resizebox{85mm}{!}{\includegraphics[angle=270]{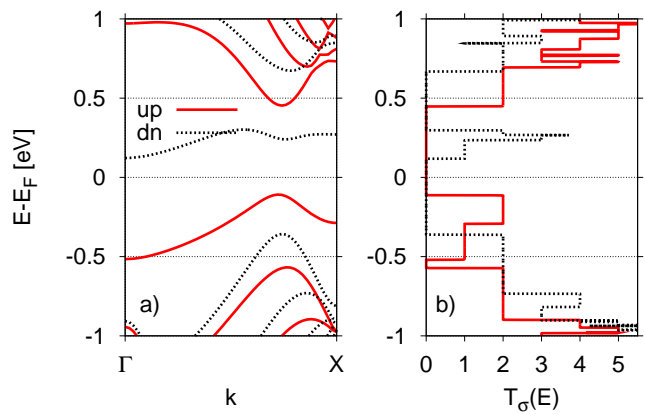}} \\
    \end{tabular}
    \caption{(Color online) Spin-resolved band structure (a) and transmission
function (b) calculated for zSiNR of 1H-2H type. The energy is measured with respect
to the Fermi energy $E_{F}$ of the corresponding pristine nanoribbon.}
    \label{fig4}
\end{figure}

\begin{figure*}[ht]
  \begin{center}
    \begin{tabular}{cc}
      \resizebox{75mm}{!}{\includegraphics[angle=270]{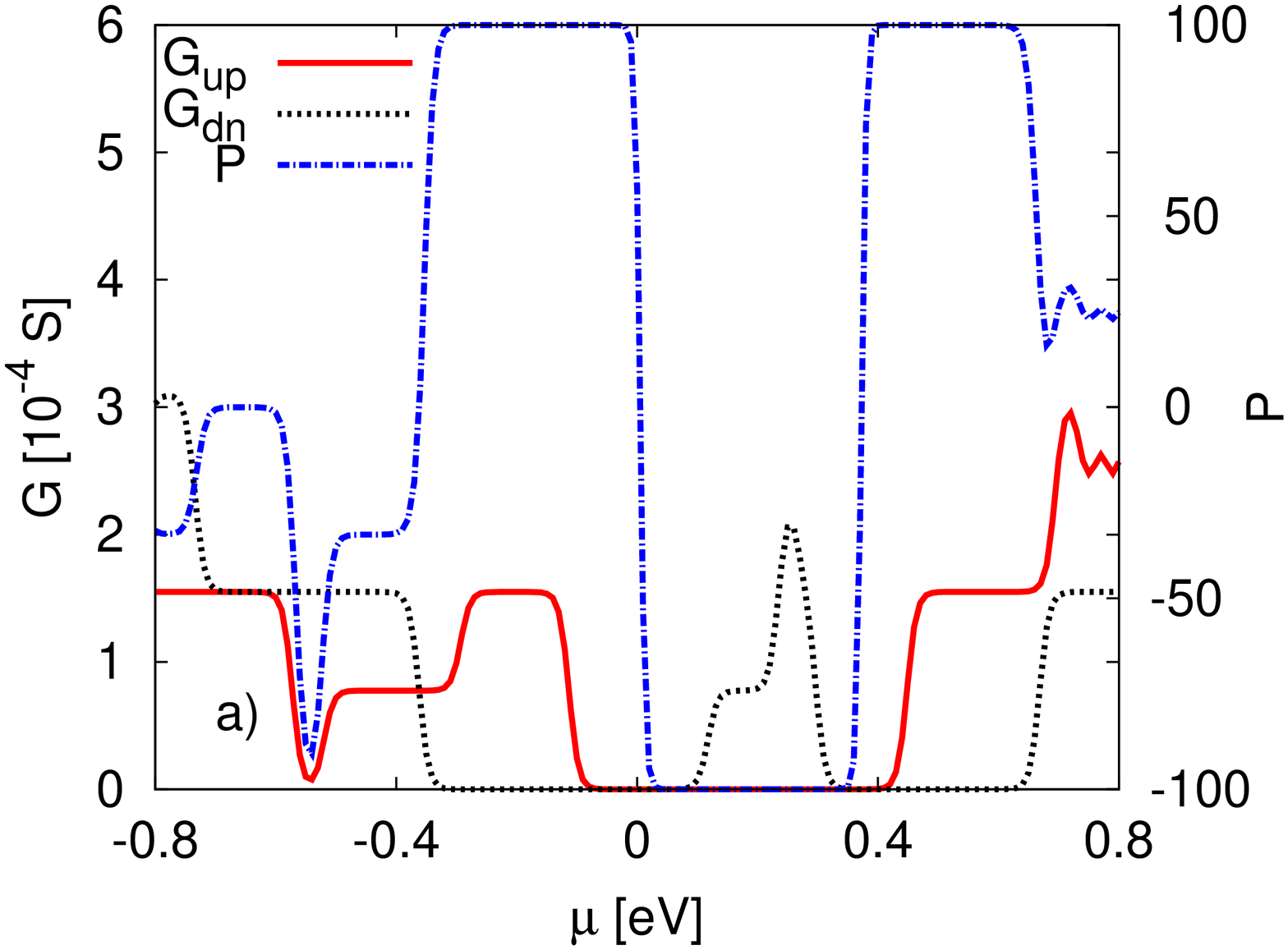}} &
      \resizebox{75mm}{!}{\includegraphics[angle=270]{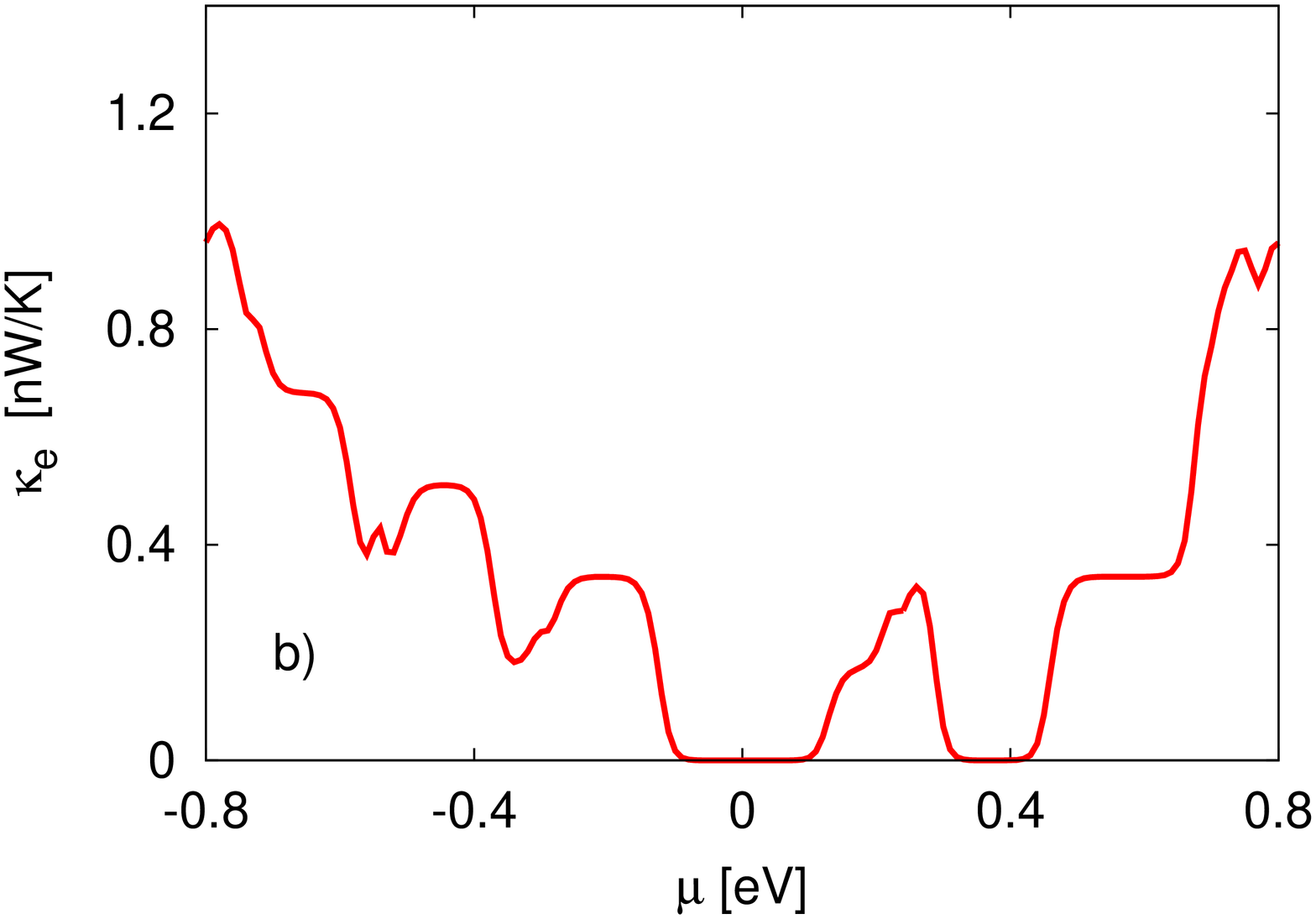}} \\
      \resizebox{75mm}{!}{\includegraphics[angle=270]{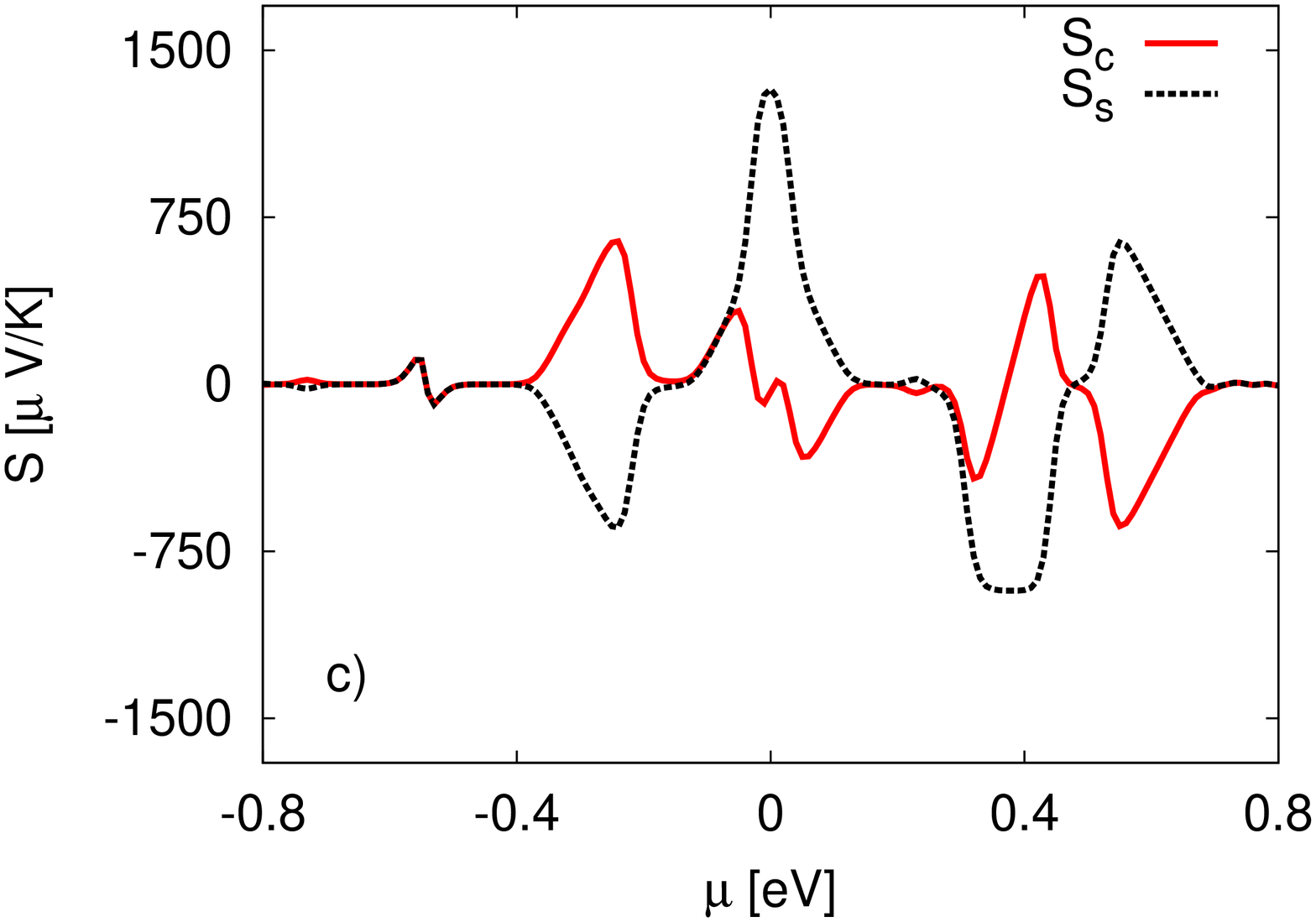}} &
      \resizebox{75mm}{!}{\includegraphics[angle=270]{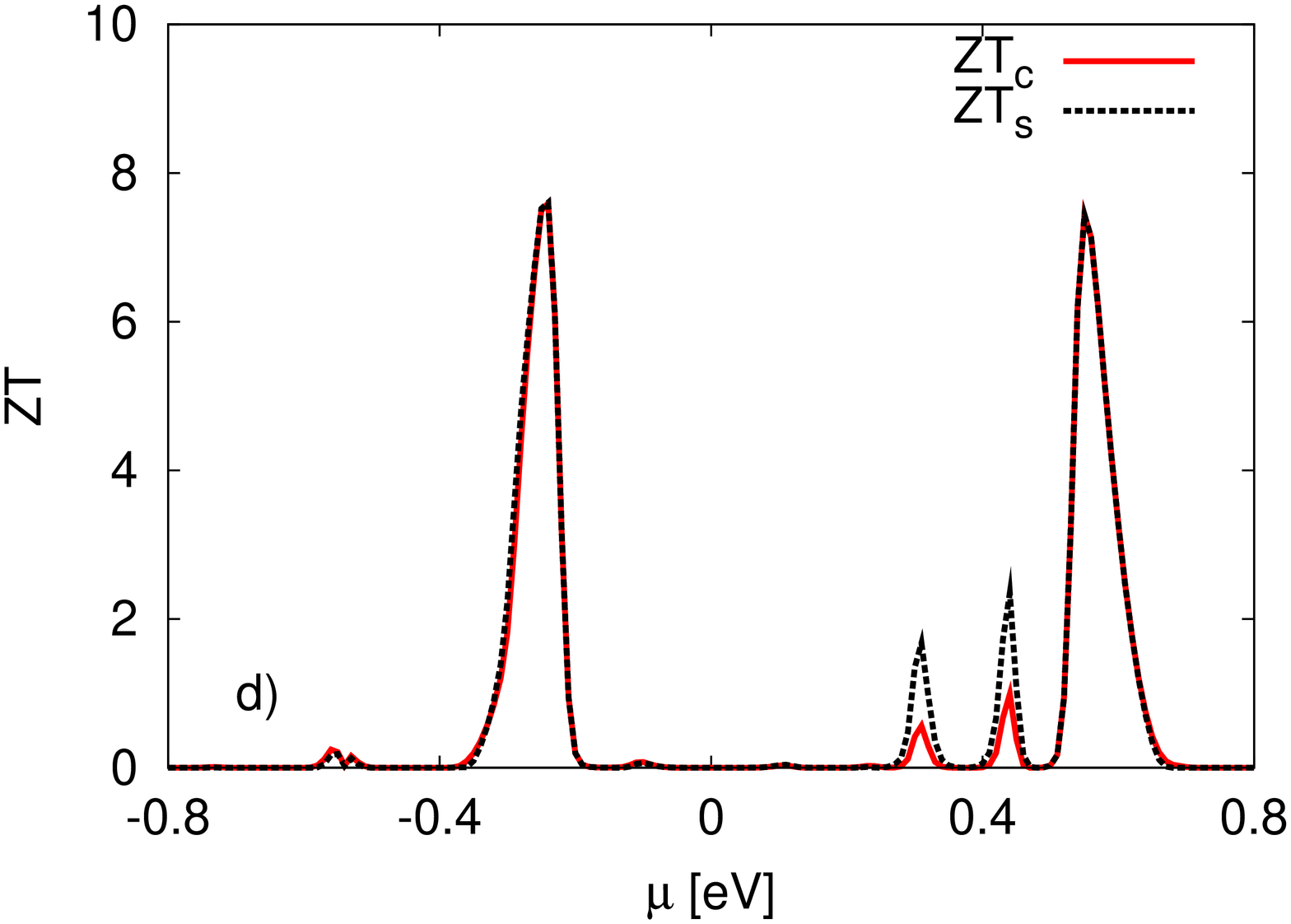}} \\
    \end{tabular}
    \caption{(Color online) Spin-resolved electric conductance $G_\sigma$ and the corresponding spin
polarization $P$ (a); electronic
thermal conductance $\kappa_{e}$ (b); charge, S$_{c}$, and spin, S$_{s}$,
thermopower (c); charge, ZT$_{c}$, and spin, ZT$_{s}$, thermoelectric
efficiency (d); calculated for zSiNR of 1H-2H type and $T=90$ K.}
    \label{fig5}
  \end{center}
\end{figure*}
Chemical potential dependence of the key transport and thermoelectric parameters is shown in Fig.5.
Similarly as in  case of 1H-0H nanoribbons, the energy gaps are strongly spin dependent. Thus, there are regions
of chemical potential where one spin channel is blocked for transport while the other channel is well conducting.
As a result, one finds in these regions almost full spin polarization of charge current, as clearly visible in Fig.5 a.
The spin polarization varies between $+100$\% and $-100$\%. Interestingly,
a small change in the chemical potential in the vicinity of the Fermi level
can results in a change of the current polarization from $P=-100$\% to $P=+100$\%.

The results obtained for electronic contribution to  the
heat conductance are now interesting as there are two energy regions of suppressed $\kappa_e$, one around $\mu = 0$
and another one for higher positive values of $\mu$. The low-temperature heat conductance in these regions is practically equal to
zero. The two regions of suppressed electronic term in the
thermal conductance  correspond to the two
true energy gaps in the electronic spectrum -- one gap for $\mu$ in the vicinity of the Fermi level and the other gap corresponding to the overlap of
the main gap in the spin-up channel and the additional gap in the spin-down channel. The second additional gap (for negative $\mu$)
is a very narrow gap in one spin channel only,
so it does not lead to  suppression of $\kappa_e$  and only contributes to the first dip seen in $\kappa_e$  for negative $\mu$.

The general behavior of the thermopower, both conventional $S_{c}$ and spin
$S_{s}$, is similar to that found in the case of in the case of 1H-0H
nanoribbons considered above,  with a huge central
maximum in $S_{s}$ near $\mu$=0. However, the side maxima of  $|S_{s}|$ and $|S_{c}|$
are shifted towards larger values of $|\mu|$. Additionally, the maximum in
$|S_{s}|$ at positive chemical potential is broad and exceeds 0.8 mV/K. This maximum
results mainly from the states near the left edge of the gap in the
spin-down channel, which gives positive contribution to $\Sdn$. Electron states
near the right edge of the spin-up gap give a negative contribution to $\Sup$.
As a result, a high value of $|S_{s}|$ can be achieved. The maximum is broad
because the peaks in $\Sup$ and $\Sdn$  are shifted with respect to each other. On
the other hand, the conventional thermopower $S_{c}$ , which corresponds to the
sum of $\Sup$ and $\Sdn$, is remarkably lower and changes sign in dependence on
which quantity, $\Sup$ or $\Sdn$, is predominant. Similar behavior can be
observed in a close vicinity of $\mu$=0, where $S_{s}$ is remarkably enhanced
while $S_{c}$ is reduced due to the opposite signs of $\Sup$ (positive) and $\Sdn$
(negative).

The conventional $ZT_{c}$ and spin $ZT_{s}$ figures of merit are presented in
Fig. 5d as a function of the chemical potential. Both types of the
thermoelectric efficiency are dominated by very high peaks appearing for
negative and positive values of $\mu$. However, due to relatively wide energy
gaps, the peaks are located at higher values of $|\mu|$, especially in the
region of positive chemical potential. The two peaks of lower intensity, which can be seen for
positive $\mu$, occur at the edges of the second energy gap in electronic spectrum
(Fig. 4). The thermopower is then considerably enhanced, especially
the spin thermopower, and the spin thermoelectric efficiency exceeds 2,
accordingly.
The results presented in this section clearly demonstrate that details of the
band structure in the vicinity of the Fermi energy strongly influence the
thermoelectric efficiency. The gaps which are required to obtain high thermopower
should be relatively narrow to give remarkable efficiency for small values of
the chemical potential.

\subsection{Antiferromagnetic nanoribbons}

Finally, let us consider nanoribbons of 2H-2H and 2H-0H types, for which the most
stable configuration is antiferromagnetic, ie. magnetic moments of the two edges
are  antiparallel. Nanoribbons of such types are semidonducting, with a wide
energy gap near the Fermi level. Here we discuss only the 2H-2H type as the
thermoelectric properties in both cases are very similar.
The band structure and transmission function for the 2H-2H nanoribbon are
presented in Fig. 6, whereas the electric and thermoelectric coefficients are
shown in Fig.7. Since the bands are spin degenerate, only the conventional thermoelectric phenomena
are relevant.
\begin{figure}[ht]
    \begin{tabular}{c}
      \resizebox{85mm}{!}{\includegraphics[angle=270]{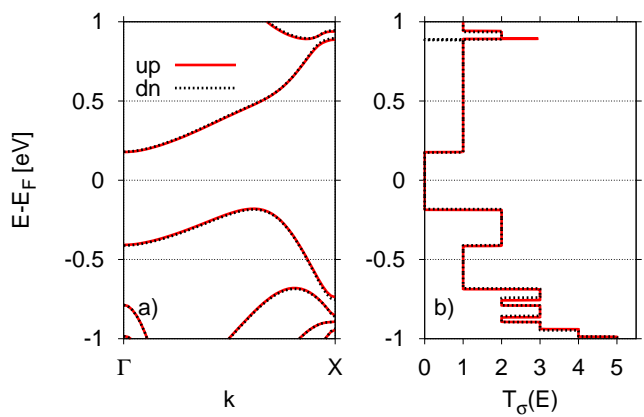}} \\
    \end{tabular}
    \caption{(Color online) Spin-resolved band structure (a) and transmission
function (b) calculated for zSiNR of 2H-2H type. The energy is measured with respect
to the Fermi energy $E_{F}$ of the corresponding pristine nanoribbon.}
    \label{fig6}
\end{figure}
\begin{figure*}[ht]
  \begin{center}
    \begin{tabular}{cc}
      \resizebox{75mm}{!}{\includegraphics[angle=270]{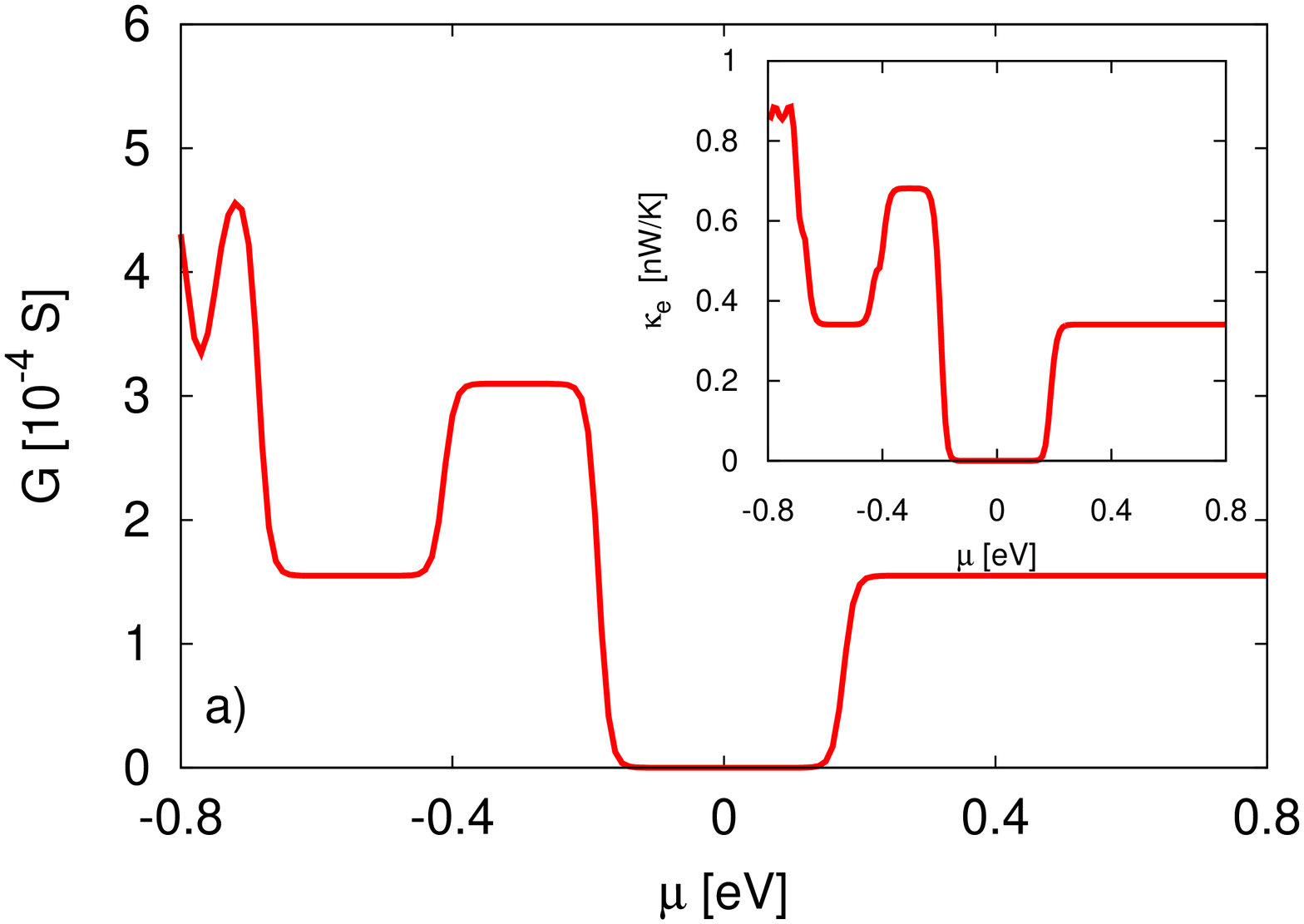}} &
      \resizebox{75mm}{!}{\includegraphics[angle=270]{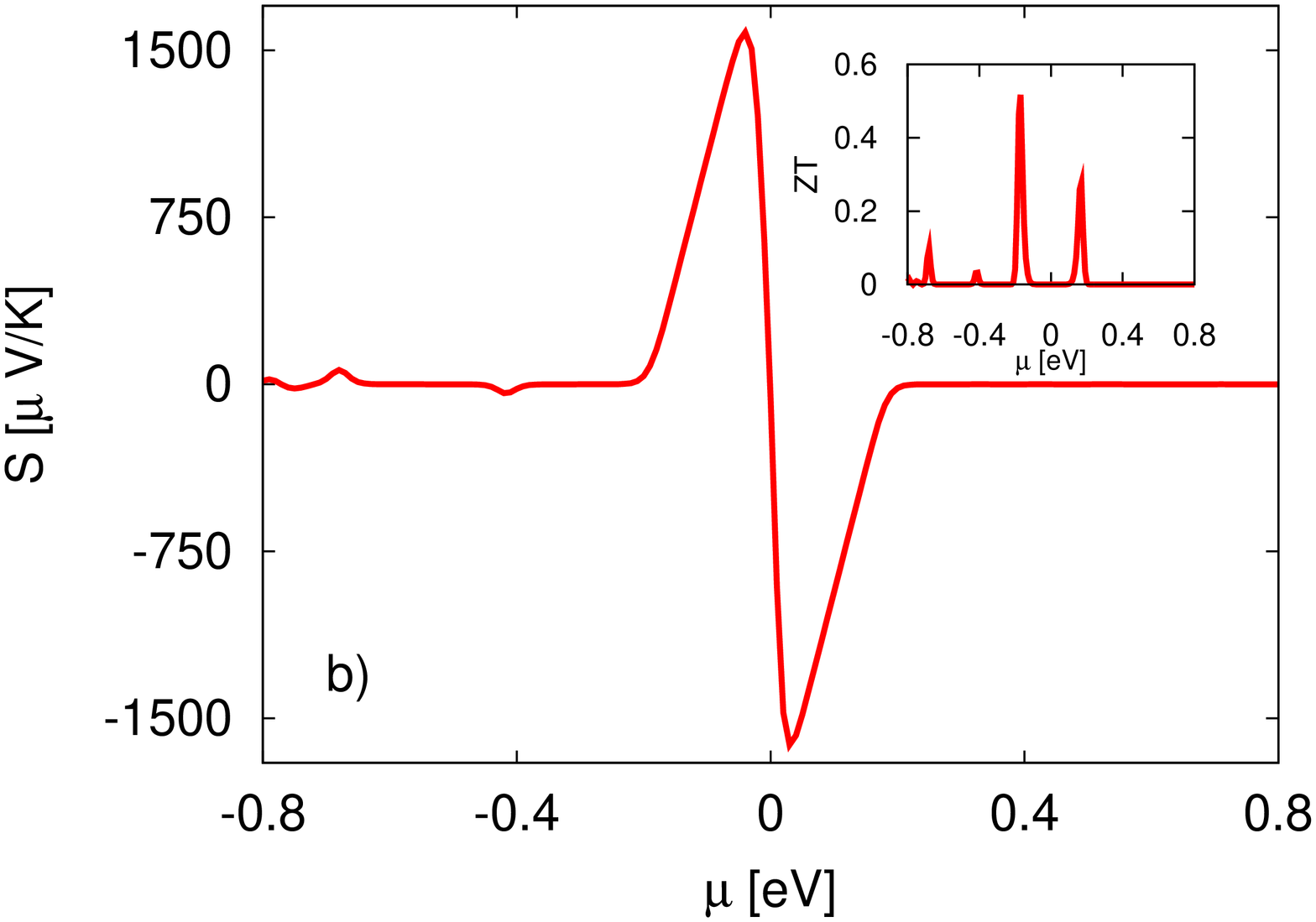}}
    \end{tabular}
    \caption{(Color online) Electric conductance per spin channel (a), electronic
thermal conductance $\kappa_{e}$ (inset to a), thermopower S (b) and thermoelectric
efficiency ZT (inset to b) calculated for zSiNR of 2H-2H type, T=90 K}
    \label{fig7}
  \end{center}
\end{figure*}

 The presence of a wide energy gap can be well visible in both electric $G$
(Fig. 7a) and electronic heat conductance $\kappa_{e}$ (inset to Fig. 7a). The
conductances increase considerably near the gap edges and are almost constant in
a wide range of chemical potential due to constant transmission. The rapid
decrease in transmission near the edge of the valence band, as well as the increase
near the edge of conductance band, give rise to a huge thermopower exceeding 1.5 mV/K (Fig.
7b). The peaks in $S$ appear at a distance of a few $kT$ from the gap edges.
However, despite of a remarkably high thermopower, the thermoelectric efficiency
$ZT$ is rather small (inset to Fig. 7b) since the electrical conductance is close to zero
in the energy gap, which essentially reduces the power factor. Similar results
are also obtained for the antiferromagnetic  nanoribbon of 2H-0H type, though
the thermoelectric efficiency in this case is lower. In general, the results are
also similar to those obtained for zSiNRs symmetrically terminated with atomic
hydrogen~\cite{b28}. They are not so spectacular as for ferromagnetic nanoribbons,
which are asymmetrically terminated with hydrogen (nanoribbons of 2H-1H and
1H-0H types discussed in the previous section). Accordingly, our analysis
clearly shows that a relatively high thermoelectric efficiency can be achieved only in systems with two non-equivalent spin channels,
in which an
energy gap can appear in one spin channel whereas the second channel supports
charge transport. Owing to this, such
systems may have a great potential for applications in spintronic and thermoelectric nanodevices.

\section {Summary and conclusions}

We have considered thermoelectric properties of zSiNRs with bare, mono- and
di-hydrogenated edges. Both symmetric as well as asymmetric cases have been
analyzed. In the former case both edges are equivalent and are either mono-or di-hydrogenated. In
the later case, in turn, the hydrogenation degree of one edge is different from
that of the other edge. The asymmetrically terminated nanoribbons of 1H-0H and
2H-1H types exhibit ferromagnetic ground state, i.e magnetic moments at the two edges are
parallel.
Such nanoribbons revel spin
thermoelectric effects in addition to the conventional ones. Moreover,
due to the specific band structure both spin and conventional thermoelectric
efficiency are remarkably enhanced.

In turn, ground-state configuration in the nanoribbons of 2H-2H and 2H-0H
types is antiferromagnetic, i.e. spin moments at one edge are opposite to those 
at the other edge.  Accordingly, both spin channels are equivalent  
and only conventional thermoelectric effects are observable.

\section{Acknowledgments}

This work was supported by the National Science Center in Poland as the Project
No. DEC-2012/04/A/ST3/00372.
Numerical calculations were performed at the Interdisciplinary Centre for
Mathematical and Computational Modelling (ICM) at Warsaw University and partly
at SPINLAB computing facility at Adam Mickiewicz University.


\end{document}